\documentclass{aa}       % New LaTeX A&A Standard Fonts
\usepackage{psfig}
\newcommand\lea{\mathrel{\raise .4ex\hbox{\rlap{$<$}\lower 1.2ex\hbox{$\sim$}}}}
\newcommand\gea{\mathrel{\raise .4ex\hbox{\rlap{$>$}\lower 1.2ex\hbox{$\sim$}}}}

\newcommand\cl{\centerline}
\renewcommand\deg{\ifmmode^\circ\else$^\circ$\fi}

\begin{document}

\thesaurus{(03.13.2; 03.13.5; 11.17.4) }

\title{Spectral distributions in compact radio sources I. Imaging with
VLBI data}

\titlerunning{Spectral imaging with VLBI data}

\author{A.P. Lobanov }

\authorrunning{A.P. Lobanov}

\offprints{A.P. Lobanov}

\institute{
Max-Planck-Institut f\"ur Radioastronomie, Auf dem H\"ugel 69, D-53121 Bonn, Germany }

\date{Received ; accepted }

\maketitle

\begin{abstract}

We discuss a technique for mapping the synchrotron turnover
frequency distribution using nearly simultaneous, multi--frequency
VLBI observations. The limitations of the technique arising from limited
spatial sampl\-ing and frequency coverage are investigated. The
errors caused by uneven spatial sampl\-ing of typical
multi--frequency VLBA datasets are estimated through numerical
simulations, and are shown to be of the order of 10\%, for pixels
with the deconvolution ${\rm SNR} \sim 7$. The fitted spectral 
parameters are corrected for the errors due to limited frequency 
coverage of VLBI data. First results from mapping the turnover frequency
distribution in \object{3C\,345} are presented.


\end{abstract}

\section{Introduction \label{sc:intro}}

Information obtained with Very Long Baseline Interferometry (VLBI)
about radio spectra of parsec--scale jets and their evolution can be
crucial for distinguishing between various jet models. However, there
are several aspects of VLBI which impede spectral studies of
parsec--scale regions.  The reliability of spectral information extracted
from VLBI data depends on many factors including sampl\-ing functions at
different frequencies, alignment of the images, calibration and
self--calibration errors, a narrow range of observing frequencies, and
source variability.  The influences of all these factors must be
understood and, if possible, corrected for, in order to reconstruct
the spectral properties of parsec--scale jets consistently.

Radio emission from the parsec--scale jets is commonly described by
the synchrotron radiation from a relativistic plasma (e.g. Pacholczyk
1970). The corresponding spectral shape, $S(\nu) \propto
\nu^{\alpha}$, is characterized by the location of spectral maximum
($S_{\rm m}$, $\nu_{\rm m}$) also called the turnover point, and by the
two spectral indices, $\alpha_{\rm thick}$ (for frequencies $\nu \ll
\nu_{\rm m}$) and $\alpha_{\rm thin}$ (for $\nu \gg \nu_{\rm m}$).

In many kiloparsec--scale objects, spectral index distributions have
been mapped, using observations made with scaled arrays. In such
observations, the antenna configurations are selected at each
frequency in a specific way such that the spatial sampl\-ings of the
resulting interferometric measurements are identical at all
frequencies used for the observations.  It is virtually impossible to
use the scaled array technique for VLBI observations of parsec--scale
jets made at different frequencies. The uneven spatial sampl\-ings of
VLBI data at different frequencies result in differences of the
corresponding synthesized beams, and can ultimately lead to confusion
and spurious features appearing in spectral index maps.

In spectral index maps, the only available kind of information is the
spectral slope between the two frequencies. While sufficient for many
purposes, this information can be misleading in the situation when the
frequency of thespectral maximum lies between the frequencies used for
spectral index mapping.  In the ranges of frequencies between 1.4 and
43\,GHz, frequently used for VLBI observations, such a situation can be
quite common. Using observations at three or more frequencies, it is
possible to estimate the shape of the synchrotron spectrum, and derive the
turnover frequency (frequency of spectral maximum). Information about
the turnover frequency can help to avoid the confusion which is
likely to occur in spectral index maps.  The turnover frequency is
sensitive to changes of physical conditions in the jet such as
velocity, particle density, and magnetic field strength. This makes it
an excellent tool for probing the physics of the jet in more detail
than is allowed by analysis of the flux and spectral index properties of
the jet.

In this paper, we present a technique suitable for determining the
turnover frequency distribution from multi--frequency VLBI data, and
investigate its limitations and ranges of applicability. We discuss
the advantages of using the Very Long Baseline Array\footnote{The Very
Long Baseline Array is operated by the National Radio Astronomy
Observatory (NRAO)} (VLBA) for spectral imaging.  A general approach
to imaging of VLBA data from nearly simultaneous, snapshot--type
observations at different frequencies is outlined in
section~\ref{sc:imaging}. The effects of limited sampl\-ing and uneven
{\it uv}--coverages are discussed in section~\ref{sc:spsens}.  We
provide analytical estimates of the sensitivity decrease, and use
numerical simulations to evaluate the effect the uneven spatial
sampl\-ings have on the outcome of a comparison of VLBI images at
different frequencies.  Alignment of VLBI images is reviewed in
section~\ref{sc:imalign}. A method used for spectral fitting and
determining the turnover frequency is described in
section~\ref{sc:fitting}. Spectral fitting in the case of limited
frequency coverage is discussed in section~\ref{sc:frcoverage}. The
first results from the turnover frequency mapping are presented in
section~\ref{sc:algorythm}.

\section{Spectral imaging using VLBA data \label{sc:imaging}}

Many of the usual technical problems in making spectral index maps
from VLBI data can be avoided by using the VLBA (see Zensus, Diamond,
\& Napier 1995). With the VLBA, it is possible to achieve array
homogeneity, have an improved flux calibration, and make the time
separation between observations at different frequencies negligible.
The major remaining problems are image alignment and uneven spatial
sampl\-ings of VLBI data taken at different frequencies.  
At an observing frequency $\nu_{\rm obs}$, the spatial sampl\-ing
of a baseline 
$B = \sqrt{B_{X}^2 + B_{Y}^2 + B_{Z}^2}$
formed by two antennas whose positions differ by $B_{X}$, $B_{Y}$,
$B_{Z}$ is characterised by the spatial frequency (e.g.
Thompson, Moran \& Swenson 1986)
\begin{equation}
\label{eq:spsens0b}
\zeta =  (\nu_{\rm obs} B /c)\sin\theta_{\rm o} \,,
\end{equation}
where $\theta_{\rm o}$ is the angle between the baseline vector and the
direction to the observed object. The quantity $\zeta$ is often
represented by its coordinate components $u$ and $v$ measured in the
spatial frequency plane ({\em uv}--plane). The spatial sampl\-ing of a VLB
array is described by the distribution of spatial frequencies
accumulated during the observation at all available baselines ({\em
uv}--coverage). In VLBI data at different frequencies, these distributions 
can differ significantly.

\begin{figure}
\cl{\psfig{figure=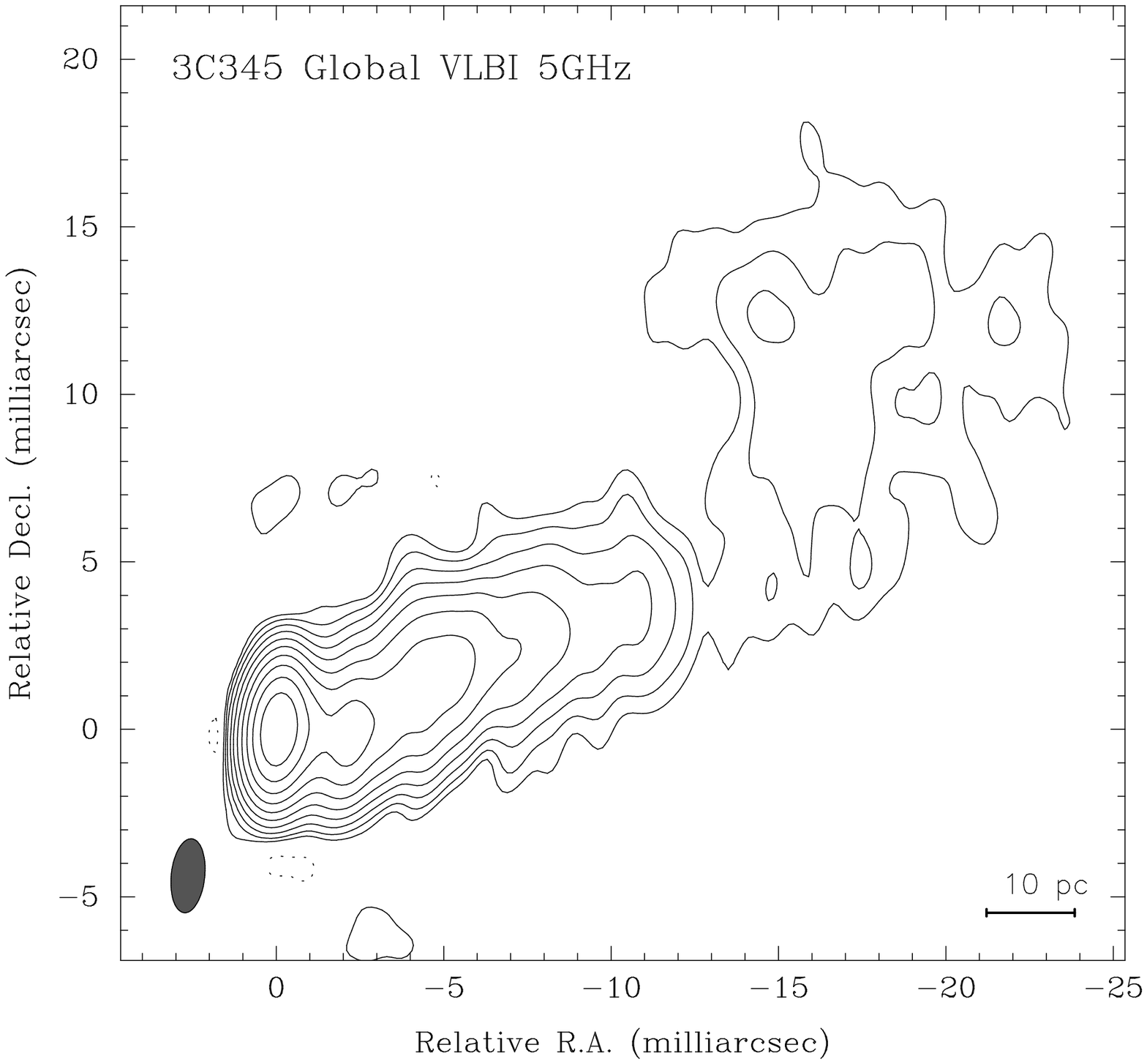,width=0.45\textwidth}}
\cl{\psfig{figure=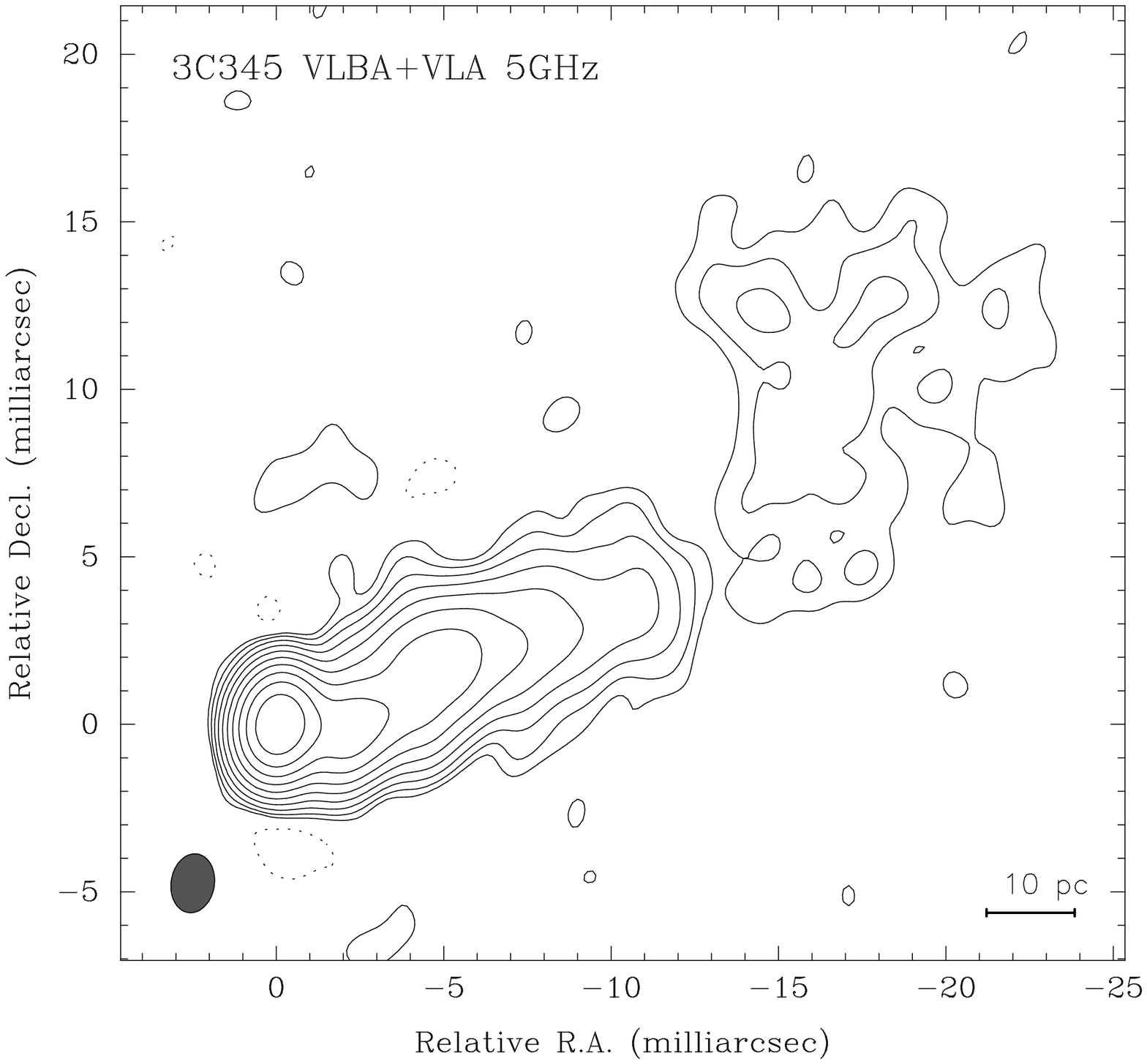,width=0.45\textwidth}}
\caption{Maps of \object{3C\,345} at 5\,GHz (Zensus et al., in prep.) obtained
from a real full--track global VLBI observation (top) and from a
simulated interleaved--frequency VLBA+VLA observation (bottom). The
contour levels are the same in both images:
$(-1,1,2,4,8,16,32,64,128,256,512)\times 3.2$\,mJy.
\label{fg:compar}}
\end{figure}

To overcome, or at least reduce, the negative effect of uneven {\em
uv}--coverages, the following scheme of observation and data
reduction can be used:

1)~Quasi--simultaneous multi--frequency observations.  The reliability
of interleaved frequency observations can be demonstrated by Figure
\ref{fg:compar} which compares a VLBI image of 3C345 obtained from a
full--track observation and an image made from simulated data that
were sampled so that they represent the {\em uv}--coverage achieved in
a 5/15 duty cycle observation (corresponding to an observation at
three frequencies, with 5 minute--long scans at each frequency).  Despite
some loss of dynamic range and sensitivity, the extended structure is
still well detected in the simulated image. The jet appearance in the
simulated image is consistent with the jet structures seen in the
actual map.

2)~Careful choice of wavelengths.  The choice of wavelengths must be a
compromise between the possibility of detecting the source structure
and the VLBA sensitivity.  Typically, at frequencies higher than
22\,GHz, the requirements on brightness temperature limit the
sensitivity.  Also, there is a stronger dependence of the high
frequency data on atmospheric instabilities.  At frequencies lower
than 2.3\,GHz, many sources will become too complicated to
warrant a successful structure detection without a full--track
observation (see Table \ref{tb:t5000}).

3)~Applying the phase--cal information for aligning the relative phases
in separate frequency bands (Cotton 1995). 

4)~Improved amplitude calibration, due to frequent system temperature
measurements and a relatively weak elevation dependence of the power
gains of the VLBA antennas (Moran \& Dhawan 1995).

\begin{table*}[t]
\caption{Maximum sampl\-ing intervals [min] and maximum detectable structure 
sizes [mas] for a 8600\,km--long baseline. 
\label{tb:t5000}}
\footnotesize
\begin{center}
\medskip
\begin{tabular}{|c|cccccccc||cccccccc|}
\hline \hline
\multicolumn{17}{c}{\bf 8600-km baseline} \\ \hline\hline
\multicolumn{1}{|c|}{ $\nu$} &
\multicolumn{8}{|c||}{\bf structure size (mas)} &
\multicolumn{8}{|c|}{\bf sampl\-ing interval (min.)} \\ 
${\rm[GHz]}$ & 1   & 3   & 5   & 10  & 20  & 30  & 50  & 100 
& 10  & 15  & 20  & 25  & 30  & 35  & 40  & 50  \\ \hline\hline

43.2& 19  & 6.5 & 4.0 & 2.0 & 1.0 & 0.5 & 0.3 & 0.2 
    & 1.9 & 1.3 & 1.0 & 0.8 & 0.7 & 0.6 & 0.5 & 0.4 \\
22.2& 36  & 12  & 7.0 & 3.5 & 1.8 & 1.2 & 0.7 & 0.4 
    & 3.5 & 2.4 & 1.8 & 1.4 & 1.2 & 1.0 & 0.9 & 0.7 \\
15.1& 55  & 18  & 11  & 5.5 & 2.7 & 1.8 & 1.1 & 0.6 
    & 5.5 & 3.8 & 2.7 & 2.2 & 1.8 & 1.6 & 1.4 & 1.1 \\
8.4 & \it 110 & 38  & 22  & 11  & 5.5 & 4.0 & 2.5 & 1.0 
    & 11  & 7.0 & 5.5 & 4.4 & 3.8 & 3.2 & 2.7 & 2.2 \\
5.0 & \it 160 & 55  & 32  & 16  & 8.0 & 5.5 & 3.0 & 1.5 
    & 17  & 11  & 8.0 & 6.5 & 5.5 & 4.6 & 4.0 & 3.2 \\
2.3 & \it 360 & \it 120 & 70  & 35  & 18  & 12  & 7.0 & 3.5 
    & 35  & 24  & 18  & 14  & 12  & 10  & 9.0 & 7.0 \\
1.6 & ... & \it 180 & \it 110 & 55  & 27  & 18  & 11  & 5.5 
    & 55  & 38  & 27  & 22  & 18  & 16  & 14  & 11  \\
0.6 & ... & ... & \it 280 & \it 140 & 70  & 46  & 28  & 14  
    & 140 & 95  & 70  & 55  & 47  & 40  & 35  & 28  \\
0.3 & ... & ... & ... & \it 240 & \it 120 & 80  & 50  & 25  
    & 240 & 160 & 120 & 100 & 80  & 70  & 60  & 50  \\
 & \multicolumn{8}{|c||}{\bf maximum sampl\-ing interval (min.)} &
   \multicolumn{8}{c|}{\bf largest detectable structure (mas)} \\
\hline
\end{tabular}
\medskip
\end{center}
\end{table*}

5)~Applying appropriate {\em uv}--tapering, in order to provide
matching {\em uv}-ranges for data taken at different frequencies.

6)~Convolving   data  at   different   frequencies   with  the   same,
artificially circular beam.

7)~Using  SNR and  flux  threshold  cutoffs,  in  order  to leave  out
the remaining sidelobe--induced artifacts.

8)~If strong sidelobe effects remain present, matching the {\em
uv}--coverages within certain {\em uv}--ranges or over the whole {\em
uv}--plane.

\section{Spatial sampl\-ing \label{sc:spsens}}

Two effects are specific to the observing scheme described above: the
decrease in sensitivity due to reduced {\it uv}--sampl\-ing at each
observing frequency, and uneven spatial sampl\-ings at different
observing frequencies. In this section, we investigate the effect 
of these factors on VLBI data at different frequencies.

 With a larger number of antennas, the overlapping parts
of the {\em uv}--coverages at different observing frequencies
constitute an increasingly larger fraction of the joint {\em
uv}--population. This results in decreasing the flux density level at
which the confusion effects dominate the results of spectral index
calculations. The confusion level can be lowered further by applying
specific weighting schemes to the data ({\em uv}--weighting), and by
matching the ranges of the {\em uv}--coverages at both frequencies.
In the spectral index maps produced using the above
procedures, confusion occurs at the flux level of about 0.3--0.5\% of
the mean peak flux density in the corresponding total intensity maps
(Lobanov 1996).

\subsection{Longest sampl\-ing intervals and largest detectable structures}

For a given time sampl\-ing interval $\Delta t$, an estimate
of the largest angular size, $\Omega_{\rm max}$, of structures that can be
detected on a baseline $B$ can be calculated from the time--average
smearing (Bridle \& Schwab 1989). The sensitivity reduction is
greatest when the apparent motion of the source is perpendicular to
the fringes associated with the selected baseline, and so we can assume
a polar source and an East--West oriented 
baseline, in order to provide the most conservative estimates of 
$\Omega_{\rm max}$. This gives
\begin{equation}
\label{eq:spsens1}
\Omega_{\rm max} = \frac{\nu_{\rm obs}}{\omega_e c \Delta t  (B_{X}^2 + 
B_{Y}^2)^{1/2}}
\,,
\end{equation}
with $\omega_e$ denoting the
Earth angular rotation speed. For a feature located at the sky coordinates
$l$, $m$ with respect to the phase--tracking center (Thompson, Moran \& 
Swenson 1986), the corresponding average reduction in amplitude over a
12--hour period is
\begin{equation}
\label{eq:spsens2}
<R_{\Delta t}> = <I/I_0> \approx 1 - \frac{\pi^2}{12 \Omega_{\rm max}^2}
(l^2 + m^2 \sin^2 \delta) \,,
\end{equation}
where $\delta$ is the source declination.

\begin{table*}
\caption{Parameters of a typical VLBA and VLA antennas\label{tb:simparam}}
\begin{center}
\footnotesize
\begin{tabular}{||rccccc|rccccc||}\hline\hline
Freq. & $T_{\rm sys}$ & K & SEFD & $\eta$ & $\sigma_{\rm therm}$ &
Freq. & $T_{\rm sys}$ & K & SEFD & $\eta$ & $\sigma_{\rm therm}$ \\
$[$GHz$]$ & [K]          & [K\,Jy$^{-1}$] & [Jy] & & [Jy]               &
$[$GHz$]$ & [K]          & [K\,Jy$^{-1}$] & [Jy] & & [Jy]               \\ \hline\hline
VLBA 0.3 &  213 & 0.100 &  2162 &  0.45 &  0.086 &
 VLA 0.3 &  150 & 0.071 &  2113 &  0.40 &  0.068 \\
     0.6 &  192 & 0.086 &  2207 &  0.40 &  0.087 &
     0.6 &  ... &   ... &  ...  &  ...  &   ...  \\
     1.6 &   29 & 0.095 &   303 &  0.57 &  0.009 &
     1.6 &   37 & 0.091 &   406 &  0.51 &  0.013 \\
     2.3 &   29 & 0.091 &   315 &  0.50 &  0.010 &
     2.3 &  ... &   ... &  ...  &  ...  &   ...  \\
     5.0 &   38 & 0.130 &   291 &  0.72 &  0.010 &
     5.0 &   44 & 0.116 &   379 &  0.65 &  0.012 \\
     8.4 &   35 & 0.117 &   304 &  0.70 &  0.009 &
     8.4 &   34 & 0.110 &   309 &  0.63 &  0.010 \\
    15.1 &   57 & 0.111 &   514 &  0.50 &  0.021 &
    15.1 &  110 & 0.093 &  1183 &  0.52 &  0.039 \\
    22.2 &   93 & 0.102 &   945 &  0.60 &  0.028 &
    22.2 &  140 & 0.082 &  1707 &  0.45 &  0.057 \\
    43.2 &  107 & 0.084 &  1348 &  0.52 &  0.038 &
    43.2 &   90 & 0.030 &  3000 &  0.37 &  0.044 \\\hline\hline
\end{tabular}
\end{center}
\end{table*}

For every
combination of wavelength and structure size, the left panel of Table
\ref{tb:t5000} gives the corresponding maximum allowed sampl\-ing interval
[in minutes] between individual scans. Dots indicate that a structure
remains unresolved. Italics highlight the 50\% decrease of
sensitivity.  The right panel of Table \ref{tb:t5000} gives the expected
maximum size of detectable structure [in mas], for all
combinations of wavelengths and sampl\-ing intervals.
The calculations have
been done for the longest
available VLBA baseline ($B_{\rm L}=8600$\,km). Here
we postulated $B_{Z}=0$ and $B_{\rm L}^2 = B_{X}^2 + B_{Y}^2$, to provide more restrictive 
estimates.

It follows from Table \ref{tb:t5000} that multi--frequency observations can
provide satisfactory structure and flux sensitivities for bright
sources with intermediate ($\approx 10$\,mas) extension.  For such
sources, full--scale spectral index mapping with 5 minute--long scans
can be done at frequencies lower than 43\,GHz
(0.7\,cm).

\subsection{Simulations of multi--frequency VLBA data}

To study the effect of uneven {\it uv}--coverages on spectral imaging,
we simulate visibility data at all frequencies available at the
VLBA, using the routine ``FAKE'' from CIT VLBI package (Pearson
1991). At all frequencies, the simulated data are produced from the
same ``CLEAN'' (Cornwell \& Braun 1989) $\delta$--component model of
the structure in the top panel of Figure \ref{fg:compar}. To improve
short--spacing coverage, we include one VLA\footnote{The Very Large Array
is operated by the National Radio Astronomy Observatory.} antenna in
the simulations.  The simulated bandwidth, $W=64$\,MHz, corresponds to
one of the standard VLBA observing modes (128\,Mb\,s$^{-1}$ data rate
with 1 bit sampl\-ing; Romney 1992). We model the antenna efficiencies,
$\eta_i$ (Crane \& Napier 1989), using the antenna sensitivities, $K_i$
(Crane \& Napier 1989), and the mean VLBA
values, $\eta_{\rm VLBA}$ and $K_{\rm VLBA}$, given in Napier (1995). Then, for
each VLBA antenna, the resulting efficiency is:
\begin{equation}
\label{eq:spsens3}
\eta_i = \eta_{\rm VLBA} (K_i/K_{\rm VLBA})\, .
\end{equation}

The system equivalent flux densities, $SEFD$, are calculated from the
antenna sensitivities and system temperatures, $T_{\rm sys}$: $SEFD_i =
T_{\rm sys,i}/K_i$ (Walker 1995; Crane \& Napier 1989).

To make the simulated data as realistic as possible, we introduce
four types of errors: the Gaussian additive noise, $\sigma_{\rm therm}$,
Gaussian multiplicative noise $\sigma_{\rm m}$, gain scaling 
errors $\sigma_{\rm gain}$, and random station--dependent phase errors. 
Both $\sigma_{\rm m}$ and $\sigma_{\rm gain}$ are chosen to be at a 2\% level,
which is a good approximation of typical VLBA gain calibration errors.

For a bandwidth of $W$[MHz] and integration time of $\tau_{\rm int}$[s], 
the additive Gaussian noise can be calculated for each antenna, 
using the antenna zenith
system temperatures, $T_{\rm sys}$, and antenna efficiencies $\eta$.
We have
\begin{equation}
\label{sc:spsens4}
\sigma_{\rm therm} = 5 T_{\rm sys} / (\eta\epsilon_{\rm pt} D_{\rm ant}^2 
\sqrt{\tau_{\rm int} W})\, ,
\end{equation}
where $D_{\rm ant} = 25$\,m is the antenna diameter, and $\epsilon_{\rm pt}$
is the pointing efficiency. The pointing efficiency can be estimated from
the ratio of pointing errors, $\sigma_{\rm pt}$, to the half--power
beamwidth of an antenna at a given frequency. The typical
non--systematic pointing errors of VLBA antennas are within
$8$--$14\arcsec$ (Romney 1992), which results in $\epsilon_{\rm pt} \sim
85$--$99\%$ at most of the VLBA observing frequencies.  The overall
parameters used in the data simulations are summarized in Table
\ref{tb:simparam} for a typical VLBA (Wrobel 1997) and VLA antennas.

\begin{figure}[h]
\cl{\psfig{figure=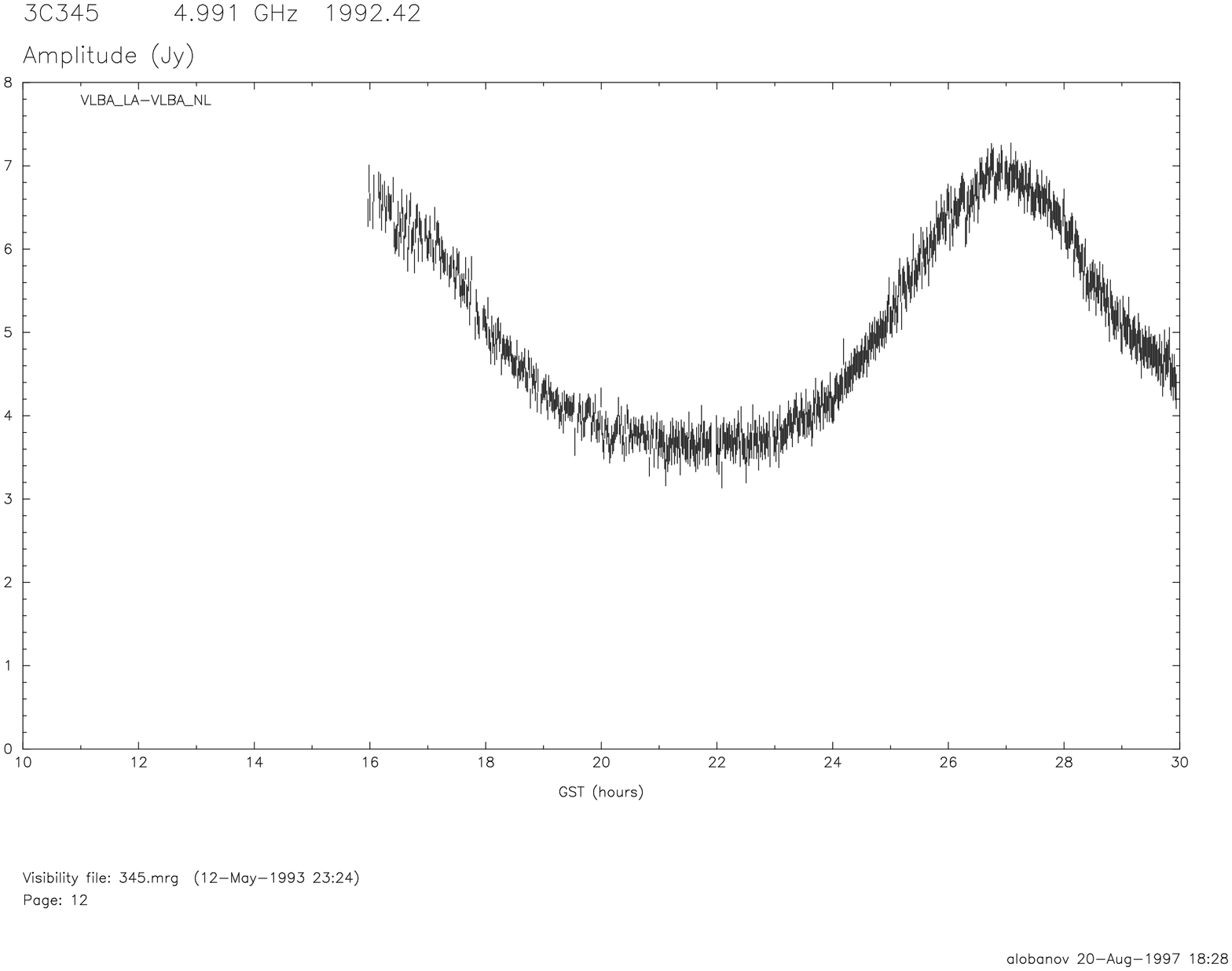,width=0.45\textwidth,height=0.12\textheight,angle=-90}}
\cl{\psfig{figure=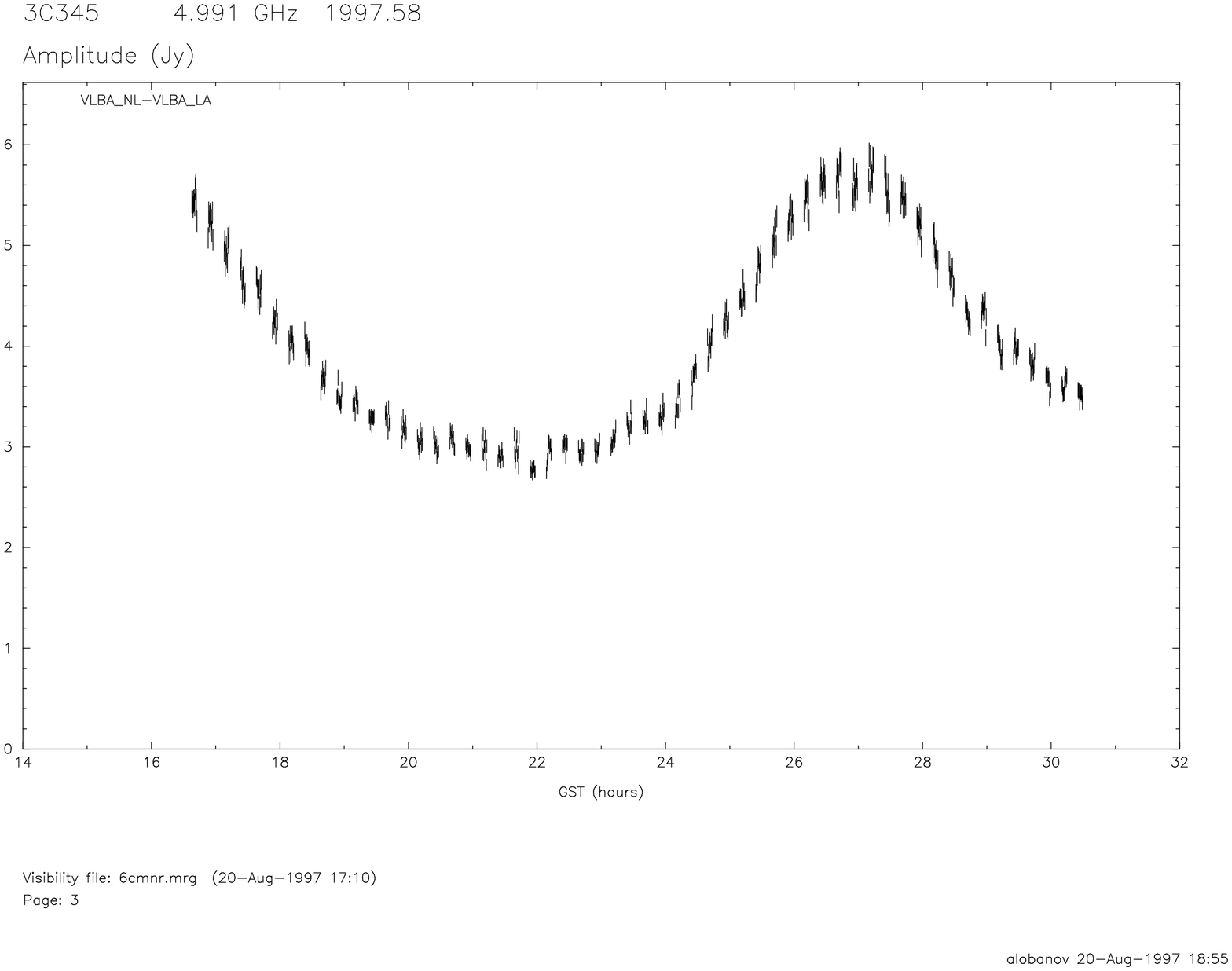,width=0.45\textwidth,height=0.12\textheight,angle=-90}}
\caption{Baseline visibility amplitudes on the baseline between VLBA 
antennas at Los Alamos and 
North Liberty. Top panel shows data from a real observation;
simulated data are shown in the bottom panel. \label{fg:baseline}}
\end{figure}

Using the parameters from Table \ref{tb:simparam}, and adding the
errors described above, we simulate VLBA visibility datasets that
would be obtained in an observation with a 5/15 duty cycle
corresponding to observing at 3 frequencies, with 5 minute long scans
at each frequency.  We simulate a 17 hour--long observation of
\object{3C\,345}, with 30 second averaging time for individual data
points---similar to the typical duration and averaging time of real VLBI
observations.  The simulated data at 5\,GHz are compared in Figure
\ref{fg:baseline} with the data from a real VLBI observation, for a
VLBA baseline Los Alamos -- North Liberty. One can see that the noise
levels are comparable in the real and simulated data.

\subsection{Spatial sampl\-ing at different frequencies}

To study the effects of uneven {\it uv}--coverages in VLBI data, we
image the simulated datasets, following the procedure described in
section \ref{sc:imaging}.  An image obtained from the simulated data
at 5\,GHz is shown in the lower panel of Figure \ref{fg:compar}. Flux
density and spectral index errors due to differences in spatial
sampl\-ings can be estimated from comparison of the images made from
the simulated data at different frequencies. In the ideal case, the
flux ratio measured between any two images should remain unity in
every pixel, and the corresponding spectral index should be zero
across the entire image. Since all images are produced from the same
source model, we ascribe all deviations from zero spectral index to
the errors due to different spatial sampl\-ings and random errors,
and choose to present these errors as a function of pixel SNR measured
with respect to the self--calibration noise. The latter is taken to be
equal to the largest negative pixel in the map, and is approximately
5--10 times bigger than the formal RMS noise of the image. In the
simulated data, the average self--calibration noise is $\approx
5$\,mJy.

\begin{figure}[t]
\cl{\psfig{figure=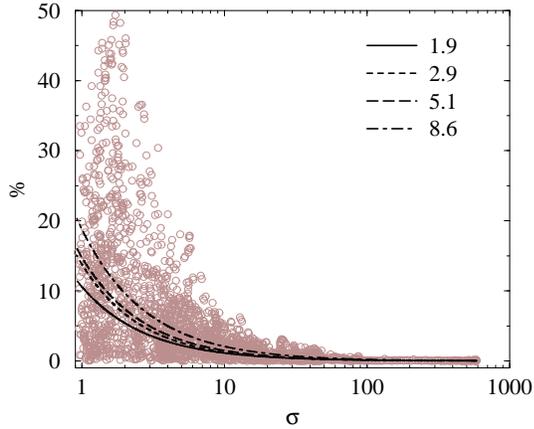,width=0.45\textwidth}}
\caption{Fractional errors due to different {\it uv}--coverages. The errors
are plotted against the pixel flux scaled to the average 
self--calibration noise, $\sigma=5$\,mJy. \label{fg:uvsens}}
\end{figure}

Figure \ref{fg:uvsens} shows the fractional errors as a function of
pixel SNR. We plot the results from all image pairs together (5, 8,
15, 22, and 43\,GHz data are used). The curved lines represent power
law fits to the individual image pairs; the frequency ratio of each
pair is given in the legend. The errors increase significantly at
SNR$\le 7$. We find that pixel SNR is the main factor determining
the derived errors, although the errors in pixels with comparable SNR
tend to increase slightly at larger distances from the phase
center. This increase however is considerably weaker compared to the
increase of errors due to lower pixel SNR.

\begin{figure}
\cl{\psfig{figure=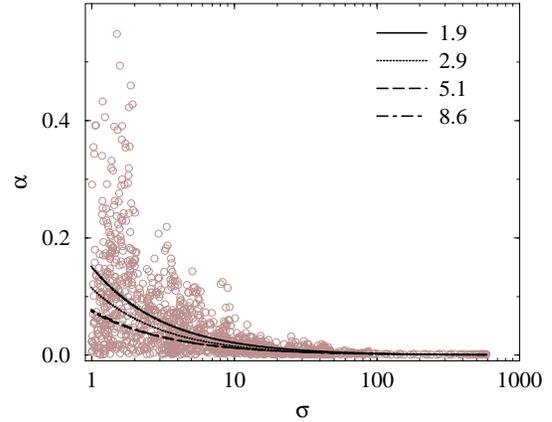,width=0.45\textwidth}}
\caption{Errors in spectral index due to different {\it uv}--coverages.
The errors are plotted against the pixel flux scaled to the average
self--calibration noise, $\sigma=5$\,mJy.
\label{fg:spixerr}}
\end{figure}

The spectral index errors are shown in Figure 
\ref{fg:spixerr} for the same image pairs. Similarly to the fractional
errors, the magnitude of spectral index errors increases rapidly at
low SNR. The main difference is that the errors become progressively
smaller at larger frequency separations, which follows obviously
from the definition of the spectral index.

\begin{figure}
\cl{\psfig{figure=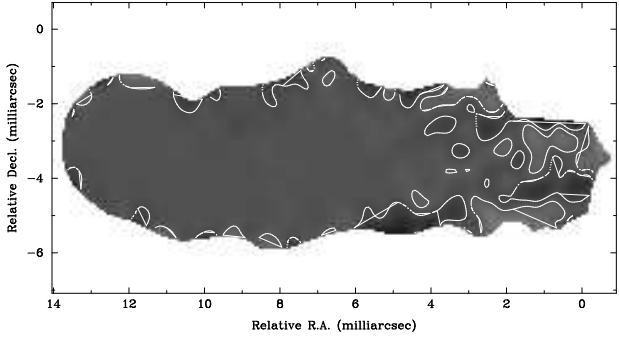,width=0.45\textwidth,angle=-90}}
\caption{Distribution of fractional errors in a 5--15\,GHz image pair.
Contours outline the regions with errors larger than 5\%. \label{fg:diverrors}}
\end{figure}

From the error distributions shown in Figures
\ref{fg:uvsens}--\ref{fg:spixerr}, we conclude that multi--frequency VLBA
observations with the time sampl\-ing interval of 10 minutes at each
frequency can be compared with each other, for pixels which are located
at moderate ($\sim10$--15\,mas) distances from the phase--tracking
center, and have a sufficiently high SNR ($\gea 5$).  Within these
limits (and with the applied observing and data reduction strategy),
the fractional errors should not exceed $\sim$10\%, a precision that
can be sufficient for several purposes including spectral index and
turnover frequency mapping in the nuclear regions of parsec--scale
jets. Most of the large amplitude errors occur at jet edges where the
effects of uneven spatial sampl\-ings are most pronounced. This
effect is readily confirmed by Figure \ref{fg:diverrors} in which the
distribution of fractional errors is plotted for the 5--15\,GHz image
pair. The contours outline areas in which the errors are larger than
5\%. These areas are concentrated at the jet edges, and cover a fairly
small fraction of the entire source structure.

\section{Image alignment \label{sc:imalign}}

Position measurements with the precision required for the alignment of
VLBI images cannot be made without extensive absolute or relative
astrometry observations.
In quasi--simultaneous multi--frequency VLBI
observations, the phase--referencing technique \cite{bc95} can be
sufficient for the purpose of image alignment. If neither of the
aforementioned techniques is available, VLBI images are usually
aligned by means of the position of compact core of the source.  The
core is likely to be located in an optically thick environment, and
its position must depend on the observing frequency, $\nu$. According
to K\"onigl (1981), the core is observed at the separation $r_{\rm
core} \propto \nu^{-1/k_r}$ from the true jet origin. The term $k_r$
is close to unity (Marcaide et al. 1985, Lobanov 1997), so that
$r_{\rm core}$ depends almost inversely on the frequency.  Therefore,
if aligned by the position of the core, VLBI images may contain
systematic position offsets undermining spectral imaging.

The frequency dependent shift of the core position can be deduced from
comparison of observations made at close epochs, assuming that the
superluminal features observed in the jet are optically thin and
therefore should have their positions unchanged. 
In this case, the offsets between the component locations measured at 
different frequencies will reflect the respective shift of the observed position of
the optically thick core. This approach has been successfully used in
several studies (e.g. Biretta et al. 1986; Zensus et al. 1995; Lobanov 1997),
and we deem it to be sufficient for the purpose of aligning VLBI images
at different frequencies.

\section{Spectral fitting \label{sc:fitting}}

\begin{figure}[t]
\cl{\psfig{figure=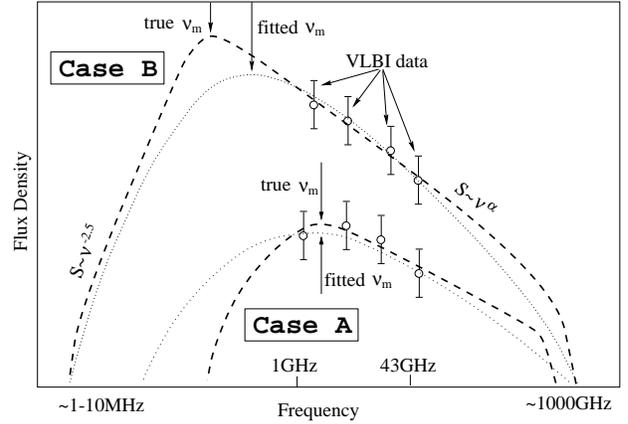,width=0.45\textwidth,angle=-90}}
\caption{Sketch illustrating the problems existing in determining the
turnover frequency from VLBI data with insufficient frequency
coverage. A homogeneous synchrotron source with isotropic pitch angle
distribution is assumed. When the true turnover frequency lies outside
of the range of observing frequencies, ad hoc information about low
and high--frequency spectral shape is required, as well as the second
order corrections based on the measured curvature of the observed part
of the spectrum (see text for details). \label{fg:tscheme}}
\end{figure}

The frequency range covered by VLBI observations is very narrow
(roughly, from 1 to 100\,GHz, with most of the observations done
between 5 and 22\,GHz). The turnover frequency, $\nu_{\rm m}$, of the 
synchrotron spectrum often lies
outside the range of observing frequencies (see the sketch in Figure 
\ref{fg:tscheme}). 
Because of the limited frequency coverage, a straightforward application
of the synchrotron spectral form to fitting may be ill--constrained 
(such as in the case B in Figure \ref{fg:tscheme}). To provide an 
estimate of $\nu_{\rm m}$ in such cases, we first attempt to
achieve the best fit of spectral data by polynomial functions, and then
analyse local curvature of the obtained fit. From the measured curvature,
an estimate (often only an upper limit) of $\nu_{\rm m}$ can
be found. Details of spectral fitting are given in section \ref{sc:erest};
application of the local curvature for determining the turnover frequency
is discussed in sections \ref{sc:frcoverage}--\ref{sc:numcor}.

\subsection{Basic synchrotron spectrum \label{sc:synbasic}}

For an extensive coverage of synchrotron emission and
its properties, we refer to the works by Ginzburg and Syrovatskii
(1969), Pacholczyk (1970), and Ternov and Mikhailin (1986).  In our
calculations, we consider synchrotron
emission from a homogeneous plasma with isotropic pitch angle
distribution and power law energy distribution
$n(\gamma)d\gamma = n_{\gamma_{0}}\gamma^{-s}d\gamma$,
for electron Lorentz factors $\gamma_{\rm L} < \gamma < \gamma_{\rm H}$.
In this case, it would suffice to describe the emission within the range of
frequencies $\nu_{\rm L} \ll \nu \ll \nu_{\rm H}$, where $\nu_{\rm L,H}$ are the 
{\em low--frequency} and {\em high--frequency} cutoffs given by:
\begin{equation}
\label{eq:syn3}
\nu_{\rm L,H} \approx \gamma^2_{\rm L,H}\frac{\Omega_e}{\pi},
\end{equation}
where $\Omega_e$ is the electron gyro--frequency.
Then, for a plasma with electron self--absorption, the spectral distribution
of emission is (Pacholczyk 1970): 
\begin{equation}
\label{eq:syn5}
I_\nu \propto \left( \frac{\nu}{\nu_1}
\right)^{\alpha_{\rm t}} \left\{ 1- \exp \left[ -\left(\frac{\nu_1}{\nu}
\right)^{\alpha_{\rm t}-\alpha_{\rm o}}\right]\right\},
\end{equation}
where $\nu_1$ is the frequency at which the optical depth,
$\tau_{\rm s}=1$, and $\alpha_{\rm t}$, $\alpha_{\rm o}$ are the spectral indices
of the optically thick and optically thin parts of the spectrum (with
spectral index defined by $S \propto \nu^\alpha$).  It is clear from
equation \ref{eq:syn5} that at frequencies $\nu \ll \nu_1$: $I_\nu
\propto (\nu/\nu_1)^{\alpha_{\rm t}}$; and at frequencies $\nu \gg \nu_1$:
$I_\nu \propto 1 - \exp [-(\nu_1/\nu)^{\alpha_{\rm t}-\alpha_{\rm o}}]$.  For
a plasma with a homogeneous synchrotron spectrum $\alpha_{\rm t}=2.5$.  We
will use the above $\alpha_{\rm t}$ and the description given by
(\ref{eq:syn5}) in our calculations.

\subsection{Fitting algorithm}

The main steps of spectral fitting can be summarized as follows:

1)~Make an approximate estimate of cutoffs in the spectrum, on the basis of
all available spectral information. A fairly good guess can be
attained by taking the average measured turnover frequency and
optically thin spectral index in the compact source. Based on these values,
calculate the high and low frequencies at which the corresponding flux
density is at an arbitrarily low level (we use $S_{\rm cutoff} =
0.1$\,mJy). Add these spectral points to the measured data in each
pixel, in order to ensure a negative curvature of the fitted curves,
as required by the theoretical spectral form (\ref{eq:syn5}).

2)~Fit the combined pixel spectra by polynomial functions, allowing
for limited variations of the cutoff frequencies, and aiming at
achieving the best fit to the measured data points. From the fits,
determine the basic spectral parameters: the turnover frequency,
turnover flux density, and integrated flux (with integration limited
to the range of observing frequencies). Estimate the errors from Monte
Carlo simulations, using the distribution of the $\chi^2$ parameters
of the fits to the simulated datasets.

3.~Calculate the local curvature of the fitted spectra within the
range of observing frequencies. Compare the derived curvature with the
values obtained from analytical or numerical calculations of the
synchrotron spectrum. Derive the corrected value of the turnover
frequency, by equating the fitted and the theoretical spectral
curvatures.

4.~Fit the data with the synchrotron spectral form described by
(\ref{eq:syn5}), and using the corrected value of the turnover frequency.

The above procedure has been applied to spectral data obtained from
modelling VLBI images of \object{3C\,345} by elliptical Gaussian components, and
combining the models at different  frequencies (Lobanov \& Zensus 1998).

\subsection{Spectral cutoffs}

Because the transition between high--frequency ($\nu > \nu_1$) and 
low--frequency
($\nu < \nu_1$) spectral regimes determined by equation \ref{eq:syn5} is 
very sharp, the
spectrum is determined by the synchrotron self--absorption at
frequencies $\nu\ll\nu_1$, and by the electron energy distribution at
$\nu\gg\nu_1$. Estimates of the typical turnover frequency and energy
spectral index in the jet of 3C345 based on our own calculations and
on the results from Raba\c{c}a and Zensus (1994) give 
$\nu_1 \approx 10$\,GHz and $s\approx 2.6$.
Using these values, we estimate the low--frequency, $\nu_{\rm L}\sim$ 1\,MHz, 
and high--frequency, $\nu_{\rm H}\sim$ 1000\,GHz, cutoffs in the spectrum,
and use these values for the spectral fitting.
However,  the cutoffs
cannot be well defined, and may change during the
evolution of the jet emission.  To account for this effect, 
we allow 15\% variations
of the cutoff frequencies so as to achieve the best fit to the data.

\subsection{Spectral fits \label{sc:erest}}

In order to calculate the spectral parameters of jet emission,
we combine the component fluxes measured at frequencies 
$\nu_1,...,\nu_{\rm N}$, and
add the cutoff information. The resulting {\em spectral dataset} 
${S_s, \sigma_s^2}$ is
represented by the flux densities,
\[
\{S_s\} =  \{ S(\nu_{\rm L}),S(\nu_1),...,S(\nu_{\rm N}),S(\nu_{\rm H})\}\,,
\]
and their respective variances,
\[
\{\sigma^2_s\} = \{ \sigma^2_S(\nu_{\rm L}),\sigma^2_S(\nu_1),...,\sigma^2_S(\nu_{\rm N}),
\sigma^2_S(\nu_{\rm H}) \}\,,
\]
where $S(\nu_{\rm L}) = S(\nu_{\rm H}) = 0.1$\,mJy is the cutoff flux density level.

By varying the spectral dataset, we then produce {\em simulated spectral 
datasets}
\begin{equation}
\label{mcf:2}
S_{\rm sym} = {\cal V}(S_s)|_{\{\sigma^2_{\rm S}(\nu)\}}
\end{equation}
assuming the Gaussian distribution of the errors in flux density measurements
\begin{equation}
\label{msf:3}
\phi (S) = \frac{1}{\sqrt{2\pi} \sigma_\nu} e^{-1/2\left(\frac{S-S_\nu}
{\sigma_\nu}\right)^2}
\end{equation}
where $S_\nu$ and $\sigma_\nu$ are the measured flux density and its variance.

For each simulated spectral dataset, we apply the $M^{th}$--order ($M\le N-1$)
polynomial fitting by the basis functions $X_k(\nu)=\nu^k$, and
minimize the $\chi^2$ parameter of the fit
\begin{equation}
\label{mcf:4}
\chi^2 = \sum^N_{i=1}\frac{1}{\sigma^2_i}\left[ (S_{\rm sym})_i - 
\sum^{M}_{k=0} A_k X_k(\nu_i) \right]^2,
\end{equation}
to obtain the least squares fit to the data, and 
derive the polynomial coefficients
\begin{equation}
\label{mcf:6}
A_j = \sum^M_{k=0} [a]^{-1}_{jk} \beta_k, \quad \sigma^2(A_j) = 
[a]{-1}_{jj} \,.
\end{equation}
Here $a_{j,k}$ and $\beta_k$ are given by
\[
a_{j,k} = \sum^N_{i=1}\frac{X_j(\nu_i)X_k(\nu_i)}{\sigma_i^2},
\]
\[
\beta_k = \sum^N_{i=1}\frac{(S_{\rm sym})_i X_k(\nu_i)}{\sigma_i^2} \, .
\]

From the fit to the spectral dataset, we derive the basic parameters of the synchrotron spectrum:
the integrated flux,
\begin{equation}
\label{mcf:7}
S_{\rm int} = \int_{\nu_1}^{\nu_{\rm N}} \sum_{j=0}^{M}A_j\nu^jd\nu \,,
\end{equation}
the turnover frequency, $\nu_{\rm m}$, and the turnover flux density, $S_{\rm m}$:
\begin{equation}
\label{mcf:8}
\frac{{\rm d}\sum_{j=0}^{M}A_j\nu^j}{{\rm d}\nu} = 0 \; \Rightarrow \; S_{\rm m}, \nu_{\rm m}
\end{equation}
We then analyse the distribution of the $\chi^2$ parameters from
the fits to all simulated datasets. From this analysis, standard 
deviations at the $3\sigma$ confidence level are calculated for $S_{\rm int}$, 
$S_{\rm m}$, and $\nu_{\rm m}$.

\subsection{Curvature of the fits \label{sc:frcoverage}}

\begin{figure}[t]
\cl{\psfig{figure=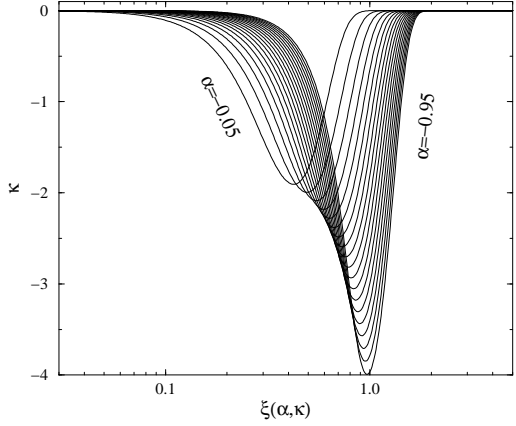,width=0.45\textwidth}}
\caption{Theoretical curvature, $\kappa$, of the homogeneous
synchrotron spectrum with spectral index $\alpha$. The $\xi$ axis
denotes the ratio of the frequency at which the curvature is
calculated to the turnover frequency. \label{fg:kappa}}
\end{figure}

\begin{figure}
\cl{\psfig{figure=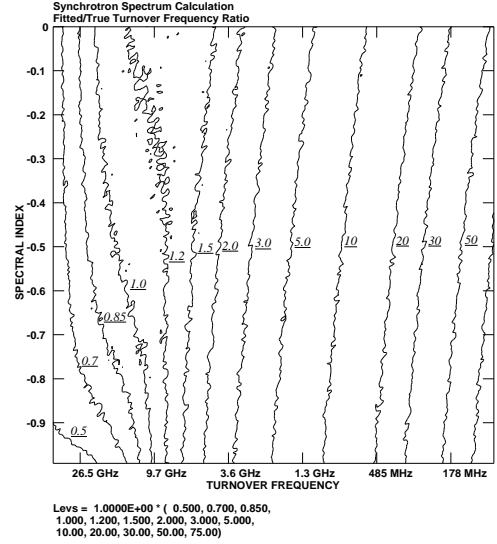,width=0.40\textwidth}}
\caption{Curvature correction coefficients derived from simulating
synchrotron spectra and subsequent fitting them by the polynomial
functions described in section 5.4. \label{fg:ratio}}
\end{figure}

The local curvature of spectral fits is:
\begin{equation}
\label{eq:korn1}
\kappa = \frac{{\rm d}^2 S}{{\rm d}\nu^2} \left[1+\left(\frac{{\rm d}S}{{\rm d}\nu}\right)^2
\right]^{-1/3}\, .
\end{equation}
The mathematical details of calculations are summarized in Appendix.
If the spectral index, $\alpha_{\rm o}$, and the local curvature,
$\kappa_{\rm o}$, of a polynomial fit are determined at a frequency
$\nu_{\rm o}$, then the turnover frequency, $\nu_{\rm m}$, can be estimated from
the adopted theoretical synchrotron spectrum. Using the derived $\alpha_{\rm o}$
and $\kappa_{\rm o}$, we determine the ratio $\xi(\alpha_{\rm o}, \kappa_{\rm o}) =
\nu_{\rm m}/\nu_{\rm o}$, from the adopted spectral form described by (\ref{eq:syn5}).
The corresponding turnover frequency is then
\begin{equation}
\label{eq:curv1}
\nu_{\rm m} = \nu_{\rm o} \xi(\alpha_{\rm o}, \kappa_{\rm o})\, .
\end{equation}
Figure \ref{fg:kappa} relates the curvature $\kappa$ of the homogeneous
synchrotron spectrum described by (\ref{eq:syn5}) to the ratio 
$\xi(\alpha_{\rm o}, \kappa_{\rm o})$. For frequencies increasingly deviating from
the turnover frequency (for which $\xi=1$), the curvature, $\kappa$,
 becomes progressively smaller, thereby limiting the ranges of applicability
of the corrections described by (\ref{eq:curv1}). For data covering the
frequencies from $\nu_{1}$ to $\nu_{\rm N}$, we expect the 
corrections to give reliable estimates for the turnover frequencies lying
within the $0.05\nu_{1} < \nu_{\rm m} < 2\nu_{\rm N}$ range.

\subsection{Numerical estimates of the curvature corrections 
\label{sc:numcor}}

An alternative method to estimate $\xi(\alpha_{\rm o}, \kappa_{\rm o})$ is to
determine it numerically, by simulating the synchrotron spectrum with
given turnover frequency and spectral index, and fitting it by the
polynomial functions used in section \ref{sc:erest}. We have performed 
such calculations, using the spectral form described in section \ref{sc:synbasic},
and covering a range of turnover frequencies and spectral indices.
The results are shown in Figure \ref{fg:ratio} in which the ratio of fitted 
values to the theoretical values of the turnover frequency is plotted against
the synchrotron spectral index. The contours show different values of
$\xi$. One can see that the determined ratios do not depend strongly on
the spectral index. Figure \ref{fg:ratio} can be used for the same purpose
of correcting the fitted turnover frequencies obtained through the procedure
described in section \ref{sc:erest}.

\begin{figure*}
\cl{\psfig{figure=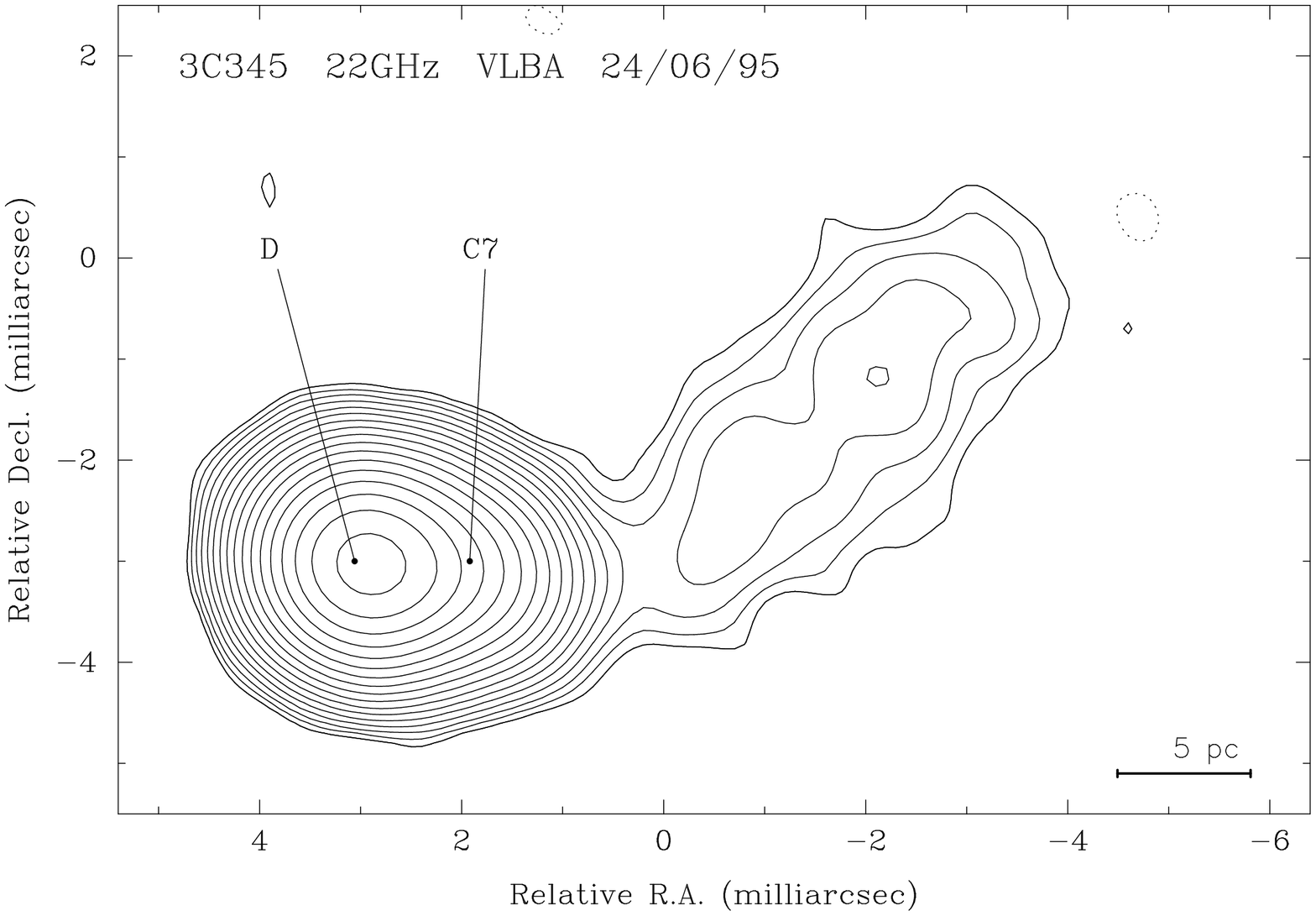,width=0.40\textwidth,angle=-90}
    \psfig{figure=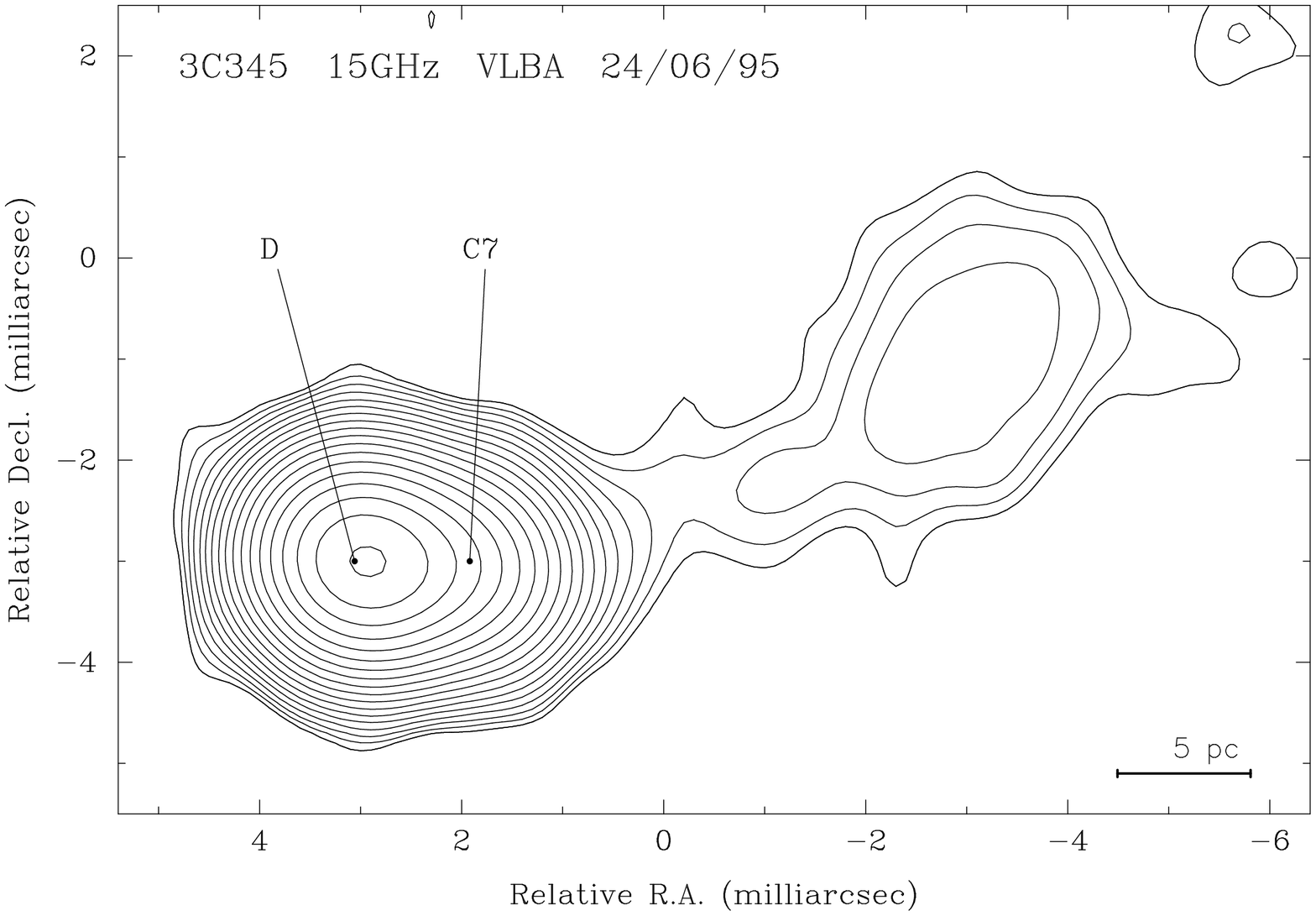,width=0.40\textwidth,angle=-90}}
\cl{\psfig{figure=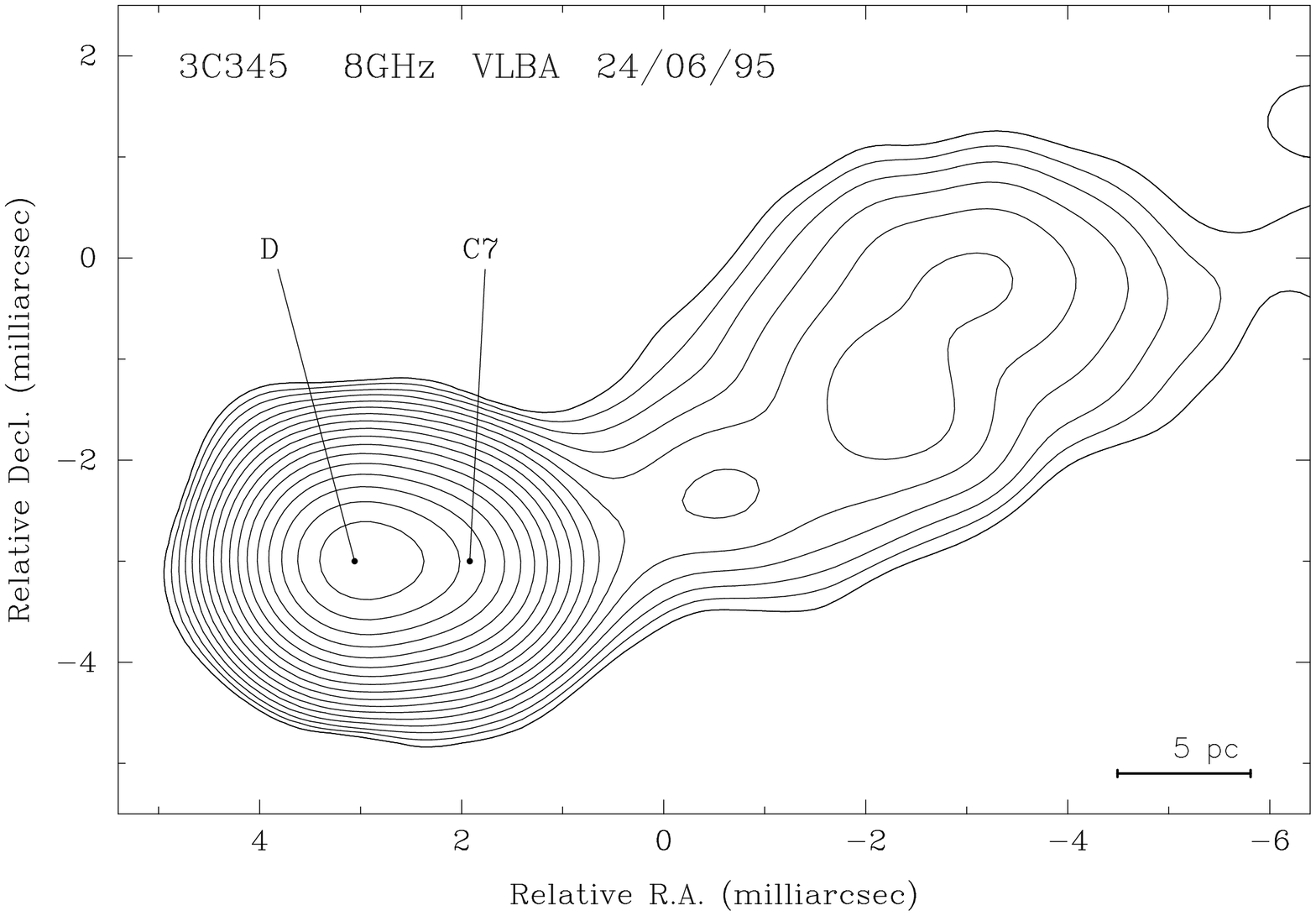,width=0.40\textwidth,angle=-90}
    \psfig{figure=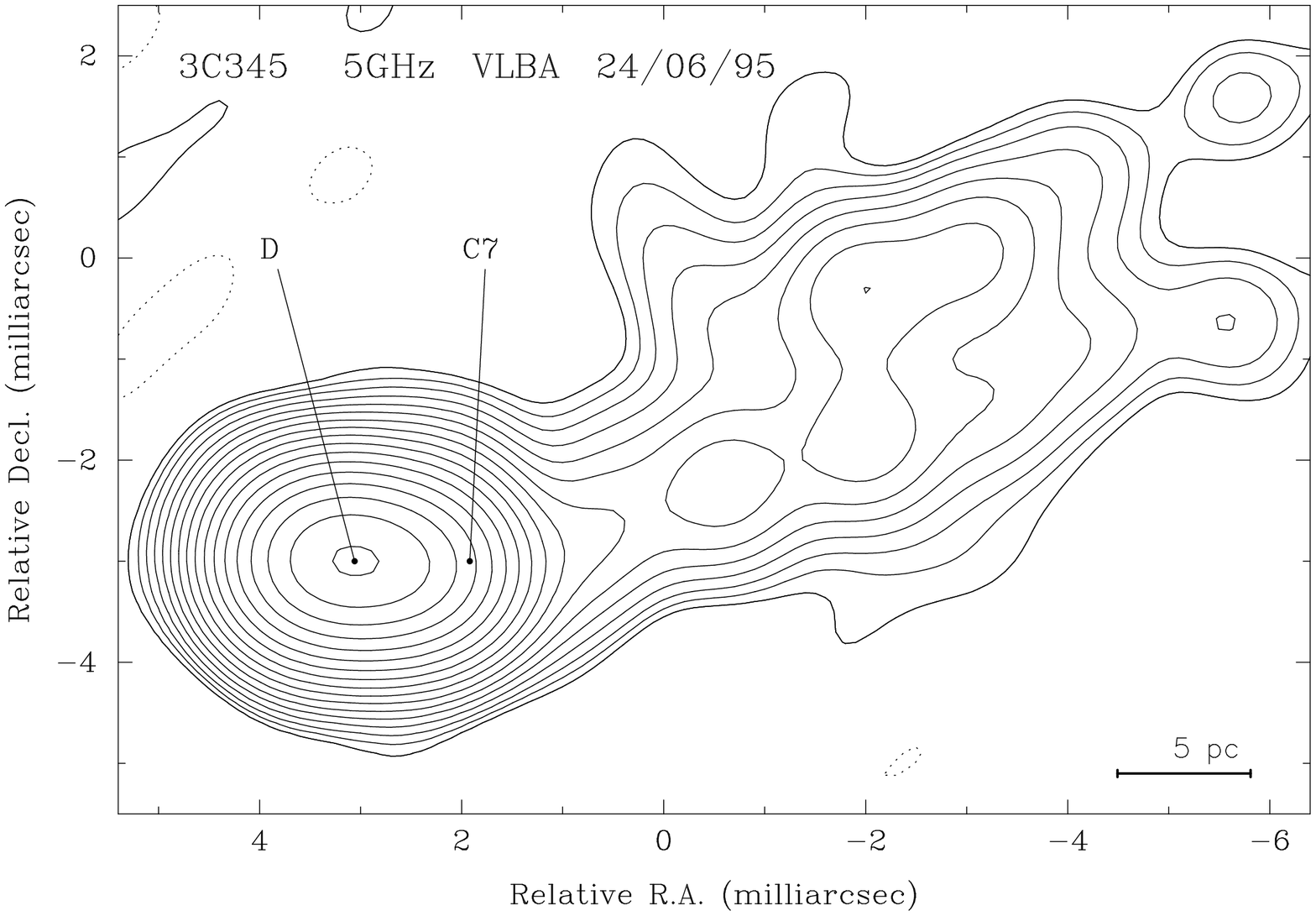,width=0.40\textwidth,angle=-90}}
\caption{Maps of \object{3C\,345} from the multi--frequency VLBA observation made 
on June 24, 1995. The restoring beam is 1.2$\times$1.2\,mas. The contours 
are (1,1.4,2,2.8,
4,...,51.2)$\times$14\,mJy.\label{fig:vlba-maps}}
\end{figure*}

\section{Turnover frequency mapping \label{sc:algorythm}}

On the basis of the method described above, we have developed a code
for mapping the turnover frequency distribution from multi--frequency
VLBA data. Fitting is performed in every valid pixel of the image; the
validation is based on clipping the pixels with low flux density or
low SNR. Flux density errors are estimated from the noise level and
flux density gradients in the total intensity maps. For each pixel, we
average the values of pixels within selected bin, and add, in
quadratures, the averaging standard deviations to the estimated noise
level. This results in slightly increased errors for pixels in the
areas with steep flux density gradients, providing more conservative
error estimates. The use of the gradients for error estimation is
optional, and can be turned off by setting the bin size to 1.

The output of the mapping procedure can be
the turnover frequency distribution, turnover flux density
distribution, integrated flux distribution, or total intensity map
at a given frequency. The last option allows us to predict, from the
fitted spectral shape, the expected source structure at any frequency
within the range of observing frequencies of the maps used for the
spectral fitting. This can also be used for testing the quality of
the spectral fit, by comparing the predicted and observed images
at the same frequency (given that the observed image was not used
for producing the above spectral fit).

\begin{table*}
\footnotesize
\begin{center}
\caption {Parameters of the VLBA maps \label{tb:vlba-map}}
\begin{tabular}{rccccccc} \hline\hline
 1 & 2& 3& 4& 5& 6& 7 &8 \\ \hline
$\nu_{\rm obs}$  & $S_{\rm tot}$ & $S_{\rm peak}$ & $S_{\rm neg}$ &
$S_{\rm noise}^{\rm est}$ & Beam & $uv$-range & $uv$-taper \\ 
$[{\rm GHz}]$ & [Jy] & [Jy/bm] & [Jy/bm] & [mJy/bm] &  & [M$\lambda$]
& [M$\lambda$] \\ \hline\hline
22.2 & 7.210 & 4.138 & -0.015 & 4.4$\pm$0.6 & 0.75$\times$0.63,
$9.^{\deg}3$ & 0-574 & 150 \\
15.4 & 7.321 & 4.439 & -0.015 & 3.9$\pm$0.4 & 0.84$\times$0.68,
$5.^{\deg}5$ & 0-440 & 150 \\
 8.4 & 7.607 & 4.173 & -0.007 & 2.4$\pm$0.2 & 1.16$\times$0.87,
$-20.^{\deg}3$ & 0-240 & 150 \\
 5.0 & 7.103 & 3.803 & -0.018 & 4.1$\pm$0.9 & 1.57$\times$1.20,
$-4.^{\deg}1$ & 0-150 & 150 \\
\hline
\end{tabular}
\medskip
\end{center}
Notes:  1 -- observing frequency; 2 -- total CLEAN flux; 3 -- peak flux density; 4 -- minimum
flux density; 5 -- measured noise; 6 -- major axis, minor axis, and
position angle of the tapered beam; 7 -- $uv$-range of the data; 8 --
half-power taper size.
\normalsize
\end{table*}

\begin{figure*}
\cl{\psfig{figure=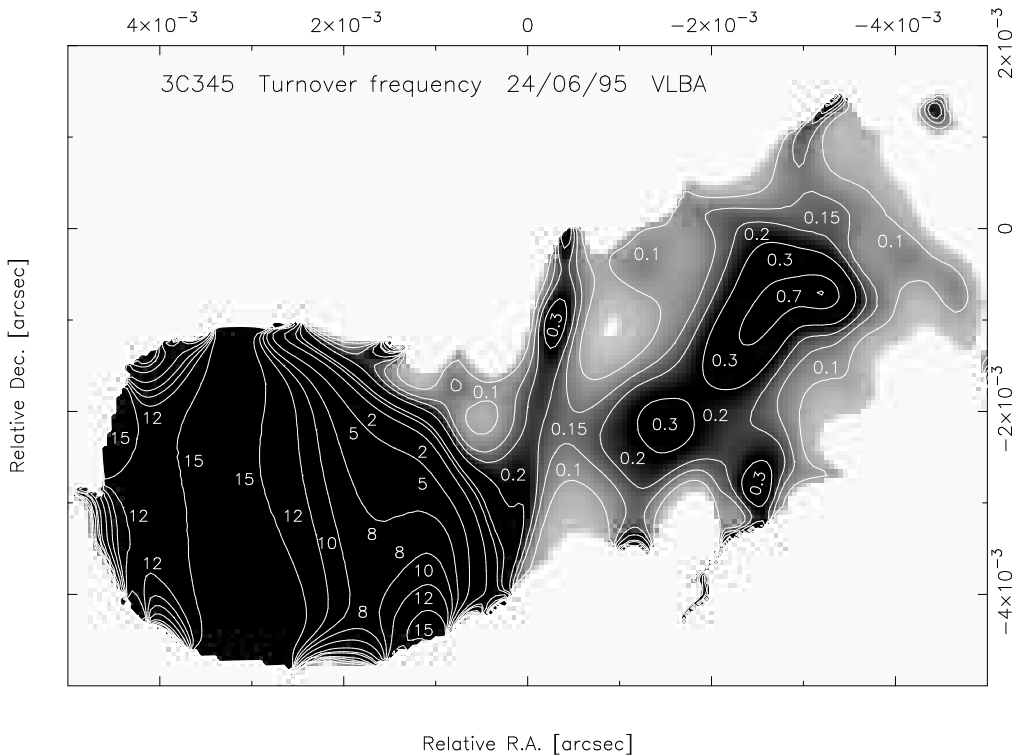,width=0.8\textwidth,angle=-90}}
\caption{Turnover frequency distribution in the extended jet of
\object{3C\,345}. The central region is saturated, for better
representation of the turnover frequency variations in the jet. The
contours are drawn at 0.1, 0.15, 0.2, 0.3, 0.7, 1, 2, 5, 8, 10, 12,
and 15\,GHz. All values below 5GHz should be regarded as upper limits.
\label{fig:tmap-jet}}
\end{figure*}

\subsection{Mapping the turnover frequency distribution in \object{3C\,345}}

The blazar \object{3C\,345} ($z=0.594$, Hewitt \& Burbidge 1993) is a strongly
variable core--jet type source with a compact core responsible for
most of the source radio emission, and a curved, parsec--scale jet
(Zensus et al. 1995) containing enhanced emission regions (bright
components) travelling along curved trajectories, with speeds of up to
20\,$c$ (Zensus et al. 1995). Synchrotron spectra of the core and the
nearest bright components are often peaked around 10\,GHz (Lobanov \&
Zensus 1998), and show a remarkable evolution. The emission from the
core and the components is believed to be produced by condensations of
highly--relativistic electron--positron plasma injected in the jet,
and losing their energy first through the inverse--Compton mechanism
(Kellermann \& Paulini-Toth 1969; Unwin et al. 1997), and later on due
to the synchrotron emission from adiabatically expanding relativistic
shocks (Wardle et al. 1994; Zensus et al. 1995).

The turnover frequency procedure was applied to the multi--frequency
VLBA observation of \object{3C\,345} made on June 24, 1995. The source was
observed at 5, 8.4, 15.4, and 22.2\,GHz. At each frequency, there was
roughly one 5 minute scan made every 20 minutes.
After the
correlation, the data were fringe--fitted and mapped in
AIPS\footnote{Astronomical Image Processing Software developed and
maintained by the National Radio Astronomy Observatory} and DIFMAP
(Shepherd 1993).  The data were tapered at 150 M$\lambda$, and the
maps were produced with a circular restoring beam of 1.2\,mas in
diameter. The core shift with respect to the reference frequency
(22.2\,GHz) was applied to the data at 5, 8.4, and 15.4\,GHz. The
magnitude of the core shift for the data at 15.4\,GHz was determined
from the fit $r_{\rm core} \propto \nu^{-1.04\pm0.16}$, whereas for the
data at 5 and 8.4\,GHz the measured values were used (Lobanov 1998).
The resulting maps are shown in Figure \ref{fig:vlba-maps}; the
main characteristics of the maps are given in Table
\ref{tb:vlba-map}. Marked in the maps are the source core ``D'' and
jet component ``C7'' which dominated the source emission at the epoch
of observation.

\begin{figure}
\cl{\psfig{figure=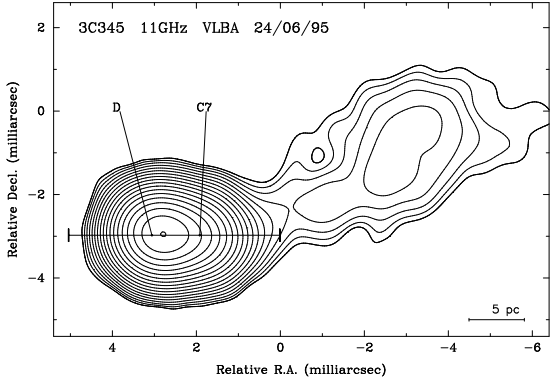,width=0.4\textwidth,angle=-90}}
\caption{Map of \object{3C\,345} at 11\,GHz obtained from the spectral fit. The 
restoring beam and contour levels are the same as in Figure 10. Spectral 
profiles in Figure 13 are taken along the horizontal
line crossing the nuclear region of the source. \label{fg:map11}}
\end{figure}

The turnover frequency map produced from the VLBA maps shown in Figure
\ref{fig:vlba-maps} is presented in Figure \ref{fig:tmap-jet}.  Figure
\ref{fg:map11} shows a map at 11\,GHz obtained from the spectral fit
to the combined VLBA data at 4 frequencies.  One can see that the main
features in the predicted 11\,GHz image are consistent with the
structures seen in the original VLBA maps.

\subsection{Nuclear region}

\begin{figure}
\cl{\psfig{figure=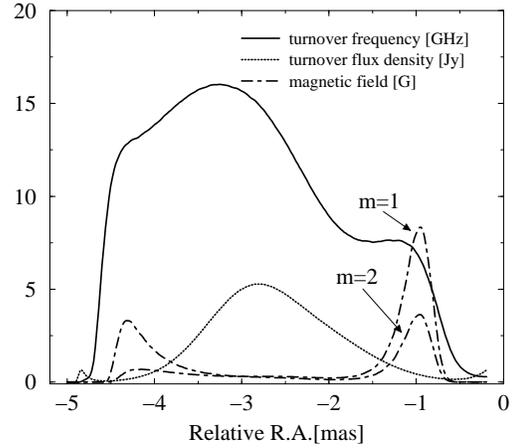,width=0.45\textwidth}}
\caption{Profiles of turnover frequency, $\nu_{\rm m}$, turnover flux,
$S_{\rm m}$, and magnetic field, $B(r)$, along the line
$\Delta\delta=-3$\,mas crossing the center of the core (horizontal
line in Figure 12). The underlying magnetic field decreases along the
jet as $r^{-m}$.\label{fig:tmap-core}}
\end{figure}

In Figure \ref{fig:tmap-jet}, there are two regions of higher turnover
frequency in the nucleus of \object{3C\,345} oriented nearly transversely to
the direction of the jet. These regions match the locations of the
core and C7 fairly well. The increased turnover frequency may indicate
that the emission is coming from a shocked plasma. The transverse
extension is then consistent with strong shocks that are likely to be
oriented almost perpendicularly to the jet direction. 
Figure \ref{fig:tmap-core} shows spectral profiles made along the horizontal
line crossing the center of the core (horizontal line in Figure 
\ref{fg:map11}. The core and C7 are both visible
in the turnover frequency profile. The turnover flux distribution is
very smooth and peaks almost precisely at the center of the core. From
the turnover frequency and turnover flux distributions, we can derive
the profile of magnetic field in the central region using the relation
(Cawthorne, 1991)
\begin{equation}
\label{sp:mf-1}
B(r) = C_0 \nu_{\rm m}^5 r^4 S_{\rm m}^{-2},
\end{equation}
where $C_0$ is the proportionality coefficient. $C_0$ can be
determined empirically from the estimates of the absolute position
($r_{\rm core}\approx 5$\,pc) and magnetic field ($B_{\rm core} \approx
0.3$\,G) of the core at 22.2\,GHz (Lobanov 1998):
\begin{equation}
C_0 = B_{\rm core} S_{\rm m,core}^2 \nu_{\rm m,core}^{-5} \approx 1.2\cdot 10^{-5}\,,
\end{equation}
for the measured $S_{\rm m,core} = 5.6$\,Jy and $\nu_{\rm m,core} = 15.1$\,GHz.
Equation \ref{sp:mf-1} expresses the magnetic field
strength due to the compression that the plasma has undergone during
shock formation. Therefore, the magnetic field also depends on the
strength of the underlying magnetic field in the location of the jet
where the shock is formed. We postulate that the underlying magnetic field
$B_{\rm amb}\propto r^{-m}$, and consider the cases, with $m=1$ and $m=2$.
The jet is assumed to have a constant opening angle 
$\phi=2.4^{\deg}$ (Lobanov 1998). 
For the magnetic field in an arbitrary pixel $p$, formula
\ref{sp:mf-1} yields 
\begin{equation}
\label{sp:mf-2}
B_{\rm p} = C_0 \nu_{\rm m,p}^5 S_{\rm m,p}^{-2} (r_{\rm p}/r_{\rm core})^{4-m}\,\,\,\,[G],
\end{equation}
In this formula, $\nu_{\rm m}$ is measured in GHz, $S_{\rm m}$ is in Jy, and $r$
is in parsecs.  The resulting magnetic field profiles are plotted in
Figure \ref{fig:tmap-core}. The magnetic field rises sharply, close to
the outer edge of C7. This can signify the amount of plasma
compression in the shock. The increased magnetic field on the opposite
side (particularly visible in the $B\propto r^{-2}$ profile) may
reflect a larger electron plasma density near the jet origin.  In the
relativistic jets, the case $B\propto r^{-1}$ is expected to be more
likely. A somewhat high value of the magnetic field in C7
($B_{\rm C7}^{\rm max} \approx 8.5$\,G) in this case may also be caused by
possible errors in the estimates of the core magnetic field. However,
the derived shape of the magnetic field profile is consistent with 
C7 being a strong shock embedded in the jet of \object{3C\,345}.

\subsection{Extended jet}

Almost everywhere in the extended jet shown in Figure
\ref{fig:tmap-jet}, the turnover frequency is lower than 5\,GHz,
posing a problem for both the spectral fitting and assessing the
results from the fits---we therefore resort to regarding all values of
$\nu_{\rm m}\le 5$\,GHz as upper limits. Apparently, there are no strong
shocks dominating the extended jet of \object{3C\,345}, or their turnover
points may have evolved rapidly due to strong adiabatic
cooling. Because the derived turnover frequencies are too low, we
cannot make quantitative statements about the physical conditions in
the extended jet. Observations at lower frequencies (1.6, 1.4, 0.6,
0.3\,GHz) are required for a better understanding of the turnover frequency
changes in these regions. With the available data, we can only make
general comments about the gradients observed in the turnover
frequency map. The bright patterns elongated along the jet ridge line
may indicate the presence of an ultra-relativistic channel inside the
jet (e.g. Sol et al 1989). The extended patterns seen in the jet at
oblique angles to the ridge line resemble the patterns of
Kelvin-Helmholtz instabilities (see Hardee et al. 1995, for the results
from 3D simulations of the KH-instability driven jets). As has been
noted above, the turnover frequency is exceptionally sensitive to the
variations of plasma speed and density.  Therefore, the observed
patterns may reflect the velocity gradients and/or density gradients
existing in the jet perturbed by the Kelvin-Helmholtz
instability. However, the low frequency data are needed for making a
better substantiated conclusion about the observed gradients.

\section{Summary \label{sc:conclus}}

In this paper, we have covered several methodological and scientific 
aspects of studying synchrotron spectrum of the parsec--scale regions in
AGN. The main conclusions can be stated as follows:

1)~We have discussed a technique that can be used for mapping the
turnover frequency distribution and obtaining spectral information
from multi--frequency VLBA data.  A feasibility study shows that
multi--frequency VLBA observations can be used for spectral imaging
and continuous spectral fitting.

2)~Multi--frequency VLBA observations made with up to 10 minute
separations between the scans at each frequency can provide a
satisfactory spatial sampl\-ing and image sensitivity for sufficiently
bright sources with intermediate ($\sim 10$--15\,mas) structures. The
fractional errors from comparing the data at different frequencies
should not exceed 10\% for emission with SNR$\ge 7$, in this case.

4)~A procedure for broadband synchrotron spectrum fitting has been
introduced for mapping the distribution of spectral parameters of
radio emission from parsec--scale jets. Corrections based on the local
curvature of the fitted spectra are introduced, in order to compensate
for the incomplete frequency coverage in cases where the true turnover
frequency is outside of the range of observing frequencies.

5)~From a 4--frequency VLBA observation of \object{3C\,345}, the first map of
the turnover frequency distribution are produced. The maps indicate
possible locations of the relativistic channel and strong shock fronts
inside the jet. The magnetic field distribution derived from the
turnover frequency and flux distributions is consistent with the plane
shocks existing in the immediate vicinity of the source core. The
extended emission appears to have a very low turnover frequency for
which the existing data do not warrant a good estimate, limiting the
conclusions to deducing certain information from the gradients of the
turnover frequency which are visible in the extended jet. The
observed gradients are consistent with the patterns of velocity
distribution and density gradients typical for Kelvin--Helmholtz
instabilities propagating in a relativistic jet. A more detailed
study, with observations made at lower frequencies, is required for
making conclusive statements about the nature of the observed
gradients of the turnover frequency.

\section*{Acknowledgements}

We would like to thank anonymous referee and I.~Pauliny-Toth for many
constructive comments on the paper.  A substantial part of this work
has been completed during the author's fellowship at the National
Radio Astronomy Observatory (NRAO).  The NRAO is a facility of the
National Science Foundation operated under cooperative agreement by
Associated Universities Inc.

\setcounter{equation}{0}
\renewcommand{\theequation}{A\arabic{equation}}
\section*{Appendix: Calculation of the local
curvature of the fitted and theoretical spectral forms}

\subsection*{Linear spectral forms}

With the fitted polynomial coefficients $a_0,...a_3$, we
can write a linear form of the fit as:
\begin{equation} 
S(\nu) = C_0 \exp(\tau)\,,
\end{equation}
with $C_0 = \exp(a_0)$ and 
\[
\tau = \sum^3_{i=1} a_i (\ln \nu)^i \, .
\]
And the derivatives used in (\ref{eq:korn1}) are given by the following
formulae:
\begin{equation}
\frac{{\rm d}S}{{\rm d}\nu} = C_0 \exp(\tau) \frac{{\rm d}\tau}{{\rm d}\nu}
\end{equation}
\begin{equation}
\frac{{\rm d}^2S}{{\rm d}{\nu}^2} = C_0 \left[\exp(\tau) \frac{{\rm d}^2\tau}{{\rm d}{\nu}^2} +
\exp(2\tau) \frac{{\rm d}\tau}{{\rm d}\nu} \right]
\end{equation}
\begin{equation}
\frac{{\rm d}\tau}{{\rm d}\nu} = \frac{1}{\nu} [a_1 + 2a_2 \ln\nu +
3a_3(\ln\nu)^2]
\end{equation}
\begin{equation}
\frac{{\rm d}^2\tau}{{\rm d}{\nu}^2} = -\frac{1}{\nu^2} [3a_3(\ln\nu)^2 +
(2a_2 - 6a_3)\ln\nu + (a_1 - 2a_2)]
\end{equation}
The power-law fit is given by:
\begin{equation}
S(\nu) = C_1 \left(\frac{\nu}{\nu_1}\right)^{\alpha_{\rm t}} \left\{ 1 -
\exp\left[-\left(\frac{\nu}{\nu_1}\right)^{\alpha_{\rm o}-\alpha_{\rm t}}\right]\right\}\,,
\end{equation}
with the corresponding derivatives:
\begin{equation}
\frac{{\rm d}S}{{\rm d}\nu} = C_1\left[\nu^{\alpha_{\rm t}} \frac{{\rm d}\,f_\nu}{{\rm d}\nu} + \alpha_{\rm t}\nu^{(\alpha_{\rm t}-1)} f_{\nu}\right] 
\end{equation}
\begin{eqnarray}
\frac{{\rm d}^2S}{{\rm d}{\nu}^2} & = & C_1[\nu^{\alpha_{\rm t}}\frac{{\rm d}^2f_{\nu}}{{\rm d}\nu^2} +
2\alpha_{\rm t}\nu^{(\alpha_{\rm t}-1)}\frac{{\rm d}\,f_{\nu}}{{\rm d}\nu} + \\ \nonumber
& & \quad\quad\quad\quad\quad\quad\quad\quad\quad\quad\quad\quad
+ (\alpha_{\rm t}-1)\alpha_{\rm t}f_{\nu}]
\end{eqnarray}
\begin{equation}
f_{\nu} = 1 - \exp\left[-\left(\frac{\nu}{\nu_1}\right)^{\lambda}\right]\,,\quad
\lambda=\alpha_{\rm o}-\alpha_{\rm t}
\end{equation}
\begin{equation}
\frac{{\rm d}\,f_{\nu}}{{\rm d}\nu} =
\frac{\lambda\nu^{\lambda-1}}{\nu_1^{\lambda}}
\,\exp\left[-\left(\frac{\nu}{\nu_1}\right)^{\lambda}\right]
\end{equation}
\begin{equation}
\frac{{\rm d}^2f_{\nu}}{{\rm d}\nu^2} = \frac{\lambda
\nu^{2\lambda-2}}{\nu_1^{2\lambda}}
[(\lambda-1)\left(\frac{\nu_1}{\nu}\right)^{\lambda} - \lambda]
\,\exp\left[-\left(\frac{\nu}{\nu_1}\right)^{\lambda}\right]
\end{equation}

\subsection*{Logarithmic spectral forms}

Following the same considerations, the logarithmic form for the polynomial fit
and its derivatives are:
\begin{equation}
S_{\rm log}(\nu) = \sum^3_{i=0} a_i(\ln\nu)^i
\end{equation}
\begin{equation}
\frac{{\rm d}S_{\rm log}}{{\rm d}(\ln\nu)} = a_1+2a_2\nu+3a_3\nu^2
\end{equation}
\begin{equation}
\frac{{\rm d}^2S_{\rm log}}{{\rm d}(\ln\nu)^2} = 2a_2+6a_3\nu
\end{equation}
And for the power-law fit:
\begin{equation}
S_{\rm log}(\nu) = \ln C_1 + \alpha_{\rm t}(\ln\nu - \ln\nu_1) + \ln f_{\nu}
\end{equation}
\begin{equation}
\frac{{\rm d}\,f_{\nu}}{{\rm d}(\ln\nu)} = \lambda
\left(\frac{\nu}{\nu_1}\right)^{\lambda}\,\exp\left[-\left(\frac{\nu}{\nu_1}\right)^{\lambda}\right]
\end{equation}
\begin{equation}
\frac{{\rm d}^2f}{{\rm d}(\ln\nu)^2} = \lambda^2 \left(\frac{\nu}{\nu_1}\right)^\lambda
\left[ 1 - \left(\frac{\nu}{\nu_1}\right)^\lambda \right]
\,\exp\left[-\left(\frac{\nu}{\nu_1}\right)^{\lambda}\right]
\end{equation}
\begin{equation}
\frac{{\rm d}S_{\rm log}}{{\rm d}(\ln\nu)} = \frac{1}{f_{\nu}}\frac{{\rm d}\,f_{\nu}}{{\rm d}(\ln\nu)} + \alpha_{\rm t}
\end{equation}
\begin{equation}
\frac{{\rm d}^2S_{\rm log}}{{\rm d}(\ln\nu)^2} = \frac{1}{f^2_{\nu}} \left[ f_{\nu}\frac{{\rm d}^2f_{\nu}}{{\rm d}(\ln\nu)^2} -
\left(\frac{{\rm d}\,f_{\nu}}{{\rm d}(\ln\nu)}\right)^2 \right]
\end{equation}

\end{document}